\documentclass[slac_one]{revtex4}
\usepackage{graphicx}
\usepackage{fancyhdr}
\usepackage{epsfig}
\newcommand{\mst}{m_{\tilde{t}_1}}
\newcommand{\neu}{\tilde{\chi}^0}
\newcommand{\mneu}[1]{m_{\tilde{\chi}^0_{#1}}}

\newcommand{\gev}{\,\, \mathrm{GeV}}

\begin{document}

\title{{\small{2005 International Linear Collider Workshop - Stanford,
U.S.A.}} 
\hfill
{\small\rm FERMILAB-CONF-05-363-E-T}\\
\vspace{12pt}
Analysis of Stops With Small Stop-Neutralino Mass Difference at a LC} 

%

\author{C.~Milst\'ene, M.~Carena, A.~Freitas}
\affiliation{Fermi National Accelerator Laboratory, Batavia, IL 60510-500, USA}
\author{A.~Finch, A.~Sopczak}
\affiliation{Lancaster University, Lancaster LA1 4YB, United Kingdom}
\author{H.~Nowak}
\affiliation{Deutsches Elektronen-Synchrotron DESY, D--15738 Zeuthen, Germany}

\begin{abstract}
A compelling framework to explain dark matter and electroweak baryogenesis is
supersymmetry with light scalar top quarks (stops) and a small mass difference
between the stop and the lightest neutralino.
In this work, the stop detection capability at the ILC for 
small mass differences between the stop and the neutralino is studied. The
analysis is based on a fast and realistic detector simulation. Significant
sensitivity for mass differences down to 5 GeV is obtained. It is discussed how
the relevant parameters of the scalar tops can be extracted and used to compute
the dark matter density in the universe.
\end{abstract}

\maketitle

\thispagestyle{fancy}


\section{INTRODUCTION} 

The origin and stabilization of electroweak symmetry, the nature of dark matter
and the generation of the baryon asymmetry (baryogenesis) in the universe
suggest the existence of new symmetries within the reach of next generation
colliders. A particular attractive extension of the Standard Model that
addresses all these issues is supersymmetry with TeV-scale supersymmetry
breaking.

A long history of experimental observations has corroborated the evidence for
dark matter in the universe, culminating in the recent accurate determination
by the WMAP satellite, in combination with the
Sloan Digital Sky Survey (SDSS) \cite{Spergel:2003cb}, $\Omega_{\rm CDM} h^2 =
0.1126^{+0.0161}_{-0.0181}$ at the 95\% C.L. Here $\Omega_{\rm CDM}$ is the
dark matter energy density normalized to the critical density and $h$ is the
Hubble parameter in units of 100 km/s/Mpc. Supersymmetry with $R$-parity
conservation provides a natural dark matter candidate, which in most scenarios
is the lightest neutralino.

Electroweak baryogenesis is based on the concept that the baryon asym\-metry 
is generated at the electroweak phase transition. While in the Standard Model
the phase transition is not sufficiently strongly first order and there is not
enough CP violation, supersymmetry can alleviate both shortcomings. A strong
first-order phase transition can be induced by loop effects of light scalar top
quarks (stops) to the finite temperature Higgs potential. In much of the
parameter space of interest for electroweak baryogenesis, the light stop is
only slightly heavier than the lightest neutralino, thus implying that
stop-neutralino co-annihilation is significant. The co-annihilation
contribution is important to meet the WMAP experimental results shown above. In
the  co-annihilation region, the stop-neutralino mass difference is smaller
than 30 GeV \cite{Balazs:2004bu}, making a discovery of the stops at hadron
colliders difficult \cite{Demina:1999ty}.

This work investigates the capabilities of a future international $e^+e^-$
linear collider (ILC) to discover light stops and measure their properties.
This study extends the reach of an earlier work \cite{stopsnew} to the region
of small stop-neutralino mass differences. The experimental analysis for the
ILC is used to assess possible conclusions about the nature of dark matter.

\section{EXPERIMENTAL SIMULATION}

In this work, the production of light stops at a 500 GeV linear collider is
analyzed, using high luminosity ${\cal L} \sim 500 {\rm \
fb}^{-1}$ and polarization of both beams.
For small mass differences $\Delta m = \mst - \mneu{1}$, a light stop decays
dominantly into a charm quark and the lightest neutralino,
$\tilde{t}_1 \to c \, \neu_1$.
The signature for stop pair production at an $e^+e^-$ collider,
\begin{equation}
e^+e^- \to \tilde{t}_1 \, \tilde{t}_1^* \to c \bar{c} \, \neu_1 \, \neu_1,
\end{equation}
is two charm jets plus missing energy. For small $\Delta m$, the jets are
relatively soft and separation from backgrounds is very challenging.
Backgrounds arising from various Standard Model processes can have
cross-sections that are several orders of magnitude larger than the signal, so
that even small jet energy smearing effects can be important. 
The possibility of efficient charm tagging plays an important role in the signal
identification and depends on the vertex detector performance. Thus it is
necessary to study this process with a realistic detector simulation.
Signal and background events are generated with {\sc Pythia 6.129}
\cite{pythia} with {\sc Circe} for realistic beamstrahlung \cite{circe}, 
together with additional private code for the stop signal 
generation previously used in Ref.~\cite{stopsnew}.  The detector simulation is
based on the fast simulation {\sc Simdet} \cite{simdet}, 
describing a typical ILC detector.

Tab.~\ref{tab:xsec} lists the cross-sections for the signal process and the
relevant backgrounds. They have been computed with the Monte-Carlo code used in
Ref.~\cite{slep} and by {\sc Grace 2.0} \cite{grace}, with cross-checks to {\sc
CompHep 4.4} \cite{comphep} where applicable. A minimal transverse momentum
cut, $p_{\rm t} > 5$ GeV, is applied for the two-photon background, to avoid
the  infrared divergence.
\begin{table}[tb]
\caption{Cross-sections for the stop signal and Standard Model background
processes for $\sqrt{s} = 500$ GeV and different polarization combinations. 
The signal is given for the stop mixing angle $\cos \theta_{\tilde{t}} = 0.5$. 
Negative/positive polarization values refer 
to left-/right-handed polarization, respectively.}
\begin{tabular}{|l|rrr|}
\hline
\bf Process &  \multicolumn{3}{c|}{\bf Cross-section [pb]} \\
\hline
$P(e^-) / P(e^+)$ & 0/0 & $\,-$80\%/+60\% & $\,$+80\%/$-$60\% \\
\hline
$\tilde{t}_1 \tilde{t}_1^* \quad$ $\mst = 120 \gev$ & 0.115 & 0.153 & 0.187 \\
\phantom{$\tilde{t}_1 \tilde{t}_1^* \quad$} $m_{\tilde{t}_1} = 140$ GeV &
  0.093 & 0.124 & 0.151 \\
\phantom{$\tilde{t}_1 \tilde{t}_1^* \quad$} $m_{\tilde{t}_1} = 180$ GeV &
  0.049 & 0.065 & 0.079 \\
\phantom{$\tilde{t}_1 \tilde{t}_1^* \quad$} $m_{\tilde{t}_1} = 220$ GeV &
  0.015 & 0.021 & 0.026 \\
\hline
$W^+W^-$ & 8.55 & 24.54 & 0.77 \\
$ZZ$	& 0.49 & 1.02 & 0.44 \\
$W e\nu$ & 6.14 & 10.57 & 1.82 \\
$e e Z$  & 7.51 & 8.49 & 6.23 \\
$q \bar{q}$, $q \neq t$ & 13.14 & 25.35 & 14.85 \\
$t \bar{t}$ & 0.55 & 1.13 & 0.50 \\
2-photon, $p_{\rm t} > 5$ GeV & 936\phantom{.00}&936\phantom{.00}&936\phantom{.00} \\
\hline
\end{tabular}
\label{tab:xsec}
\end{table}

In the first step of the event selection, the following cuts are applied:
\begin{equation}
\begin{array}{lll}
4 < N_{\mbox{\scriptsize charged tracks}} < 50, &\quad &p_{\rm t} > 5 \gev, \\
|\cos \theta_{\rm thrust}| < 0.8, & & |p_{\rm long,tot}/p_{\rm tot}| < 0.9, \\
E_{\rm vis} < 0.75 \sqrt{s}, & & m_{\rm inv} < 200 \gev.
\end{array}
\end{equation}
The cut on the number of charged tracks removes most leptonic background 
and part of the $t\bar{t}$ background. By requiring a
minimal transverse momentum $p_{\rm t}$, the two-photon background and
back-to-back processes like $q\bar{q}$ are largely reduced. 
The signal is characterized by large missing energy and transverse momentum
from the two neutralinos, whereas for most backgrounds the missing momentum
occurs from particles lost in the beam pipe. Therefore, cuts on the thrust
angle $\theta_{\rm thrust}$, the longitudinal momentum $p_{\rm long,tot}$, the
visible energy $E_{\rm vis}$ and the total invariant mass $m_{\rm inv}$ are
effective on all backgrounds.
\begin{table}[tb]
\caption{Background event numbers and $\tilde{t}_1 \tilde{t}_1^*$
signal efficiencies in \% (for various $\mst$ and $\Delta m$ in GeV) 
after preselection
and each of the final selection cuts. In the last column the expected event
number are scaled to a luminosity of 500~fb$^{-1}$.
The cuts are explained in the text.}
\begin{tabular}{|ll|r|r|rrrrrr|r|}
\hline
 & & & After$\;$\rule{0mm}{0mm} &  & &  &  &  & 	& Scaled to \\[-.5ex]
Process & & Total & presel. & cut 1 & cut 2 & cut 3 & cut 4 & cut 5 & cut 6
	& 500 fb$^{-1}$ \\
\hline
&$W^+W^-$ & 210,000 & 2814 & 827 & 28 & 25 & 14 & 14 & 8 & 145 \\ 
&$ZZ$	& 30,000 & 2681 & 1987 & 170 & 154 & 108 & 108 & 35 & 257 \\ 
&$W e\nu$ & 210,000 & 53314 & 38616 & 4548 & 3787 & 1763 & 1743 & 345 & 5044 \\ 
&$e e Z$  & 210,000 & 51 & 24 & 20 & 11 & 6 & 3 & 2 & 36 \\ 
&$q \bar{q}$, $q \neq t$ & 350,000 & 341 & 51 & 32 & 19 & 13 & 10 & 8 & 160 \\ 
&$t \bar{t}$ & 180,000 & 2163 & 72 & 40 & 32 & 26 & 26 & 25 & 38 \\
&2-photon & $8 \times 10^6$ & 4061 & 3125 & 3096 & 533 & 402 & 0 & 0 & $<164$\\
\hline
$\mst = 140$ &
\rule{0mm}{0mm} $\; \Delta m = 20$ & 50,000 & 68.5 & 48.8 & 42.1 & 33.4 &
	27.9 & 27.3 & 20.9 & 9720 \\ 
&\rule{0mm}{0mm} $\; \Delta m = 40$ & 50,000 & 71.8 & 47.0 & 40.2 & 30.3 &
	24.5 & 24.4 & 10.1 & 4700 \\ 
&\rule{0mm}{0mm} $\; \Delta m = 80$ & 50,000 & 51.8 & 34.0 & 23.6 & 20.1 &
	16.4 & 16.4 & 10.4 & 4840 \\ 
$\mst = 180$ &
\rule{0mm}{0mm} $\; \Delta m = 20$ & 25,000 & 68.0 & 51.4 & 49.4 & 42.4 &
	36.5 & 34.9 & 28.4 & 6960 \\ 
&\rule{0mm}{0mm} $\; \Delta m = 40$ & 25,000 & 72.7 & 50.7 & 42.4 & 35.5 &
	28.5 & 28.4 & 20.1 & 4925 \\ 
&\rule{0mm}{0mm} $\; \Delta m = 80$ & 25,000 & 63.3 & 43.0 & 33.4 & 29.6 &
	23.9 & 23.9 & 15.0 & 3675 \\ 
$\mst = 220$ &
\rule{0mm}{0mm} $\; \Delta m = 20$ & 10,000 & 66.2 & 53.5 & 53.5 & 48.5 &
	42.8 & 39.9 & 34.6 & 2600 \\ 
&\rule{0mm}{0mm} $\; \Delta m = 40$ & 10,000 & 72.5 & 55.3 & 47.0 & 42.9 &
	34.3 & 34.2 & 24.2 & 1815 \\ 
&\rule{0mm}{0mm} $\; \Delta m = 80$ & 10,000 & 73.1 & 51.6 & 42.7 & 37.9 &
	30.3 & 30.3 & 18.8 & 1410 \\ 
\hline
\end{tabular}
\label{tab:evtn}
\end{table}
As can be seen in Tab.~\ref{tab:evtn}, the various background are substantially
reduced after these preselection cuts, while roughly around 70\% of the signal
is preserved.

After generating large event samples with the preselection cuts
for the various backgrounds as listed in Tab.~\ref{tab:evtn},
the following final event selection cuts are applied to further improve the
signal-to-background ratio:
\begin{enumerate}

\item Number of jets $N_{\rm jets} = 2$. Jets are
reconstructed with the Durham algorithm with the jet resolution parameter
$y_{\rm cut} = 0.003 \times \sqrt{s}/E_{\rm vis}$. The cut
reduces substantially the number of $W$ and quark pair events.

\item Large missing energy, $E_{\rm vis} < 0.4 \sqrt{s}$. This is
effective against $W^+W^-$, $ZZ$ and di-quark events. 
In addition, a window for the invariant jet mass around the
$W$-boson mass, 
$70 \gev < m_{\rm jet,inv} < 90 \gev$,
is excluded to reduce the large $W e \nu$ background.

\item $q\bar{q}$ events are removed by requiring a minimal acollinearity angle  $\cos \phi_{\rm
aco} > -0.9$. 

\item Cutting on the thrust angle, $|\cos \theta_{\rm thrust} < 0.7|$, reduces
$W$ boson background.

\item A strong cut on the transverse momentum, $p_{\rm t} > 12 \gev$,
completely removes the remaining two-photon events.

\item The largest remaining background is from $e^+e^- \to W e\nu$. It
resembles the signal closely in most distributions, {\it e.g.} as a function of
the visible energy, thrust or acollinearity. Only by increasing the invariant
jet mass window from cut 2 to ($60 \gev < m_{\rm jet,inv} < 90 \gev$), the
signal-to-background ratio is improved, but at the cost of losing a substantial
amount of the signal. In addition, the signal selection can be enhanced by
charm tagging, which is implemented based on the neural network analysis
described in Ref.~\cite{kuhlhiggs}. The neural network has been optimized to
reduce the $We\nu$ background while preserving the stop signal for small mass
differences. 

\end{enumerate}

The resulting event numbers, scaled to a luminosity of 500~fb$^{-1}$, and the
signal efficiencies are listed in Tab.~\ref{tab:evtn}. After the final
selection, the $\tilde{t}_1 \tilde{t}_1^*$ signal event numbers are of
same order as the remaining background, $N \sim {\cal O}(10^4)$.

To explore the reach for very small mass
differences $\Delta m = \mst - \mneu{1}$, signal event samples have been
generated also for  $\Delta m = 10 \gev$ and 5 GeV, see
Tab.~\ref{tab:sigeff}.
\begin{table}[tb]
\caption{Signal efficiencies for $\tilde{t}_1 \tilde{t}_1^*$ production after
final event selection for different combinations of the stop mass $\mst$ and
mass difference $\Delta m = \mst - \mneu{1}$.}
\begin{tabular}{|r|r@{}cccc|}
\hline
$\Delta m$ & $m_{\tilde{t}_1} =\;$ & 120 GeV & 140 GeV & 180 GeV & 220 GeV \\
\hline
80 GeV && & 10\% & 15\% & 19\% \\
40 GeV && & 10\% & 20\% & 24\% \\
20 GeV && 17\% & 21\% & 28\% & 35\% \\
10 GeV && 19\% & 20\% & 19\% & 35\% \\
5 GeV &&  2.5\% & 1.1\% & 0.3\% & 0.1\% \\
\hline
\end{tabular}
\label{tab:sigeff}
\end{table}
The signal efficiency drastically drops for $\Delta m
= 5 \gev$, as a result of the $p_{\rm t}$ cut (cut 5). An optimization of the
event selection for very small $\Delta m$ will be addressed in future
work.

\begin{figure}[tb]
\centering
\epsfig{file=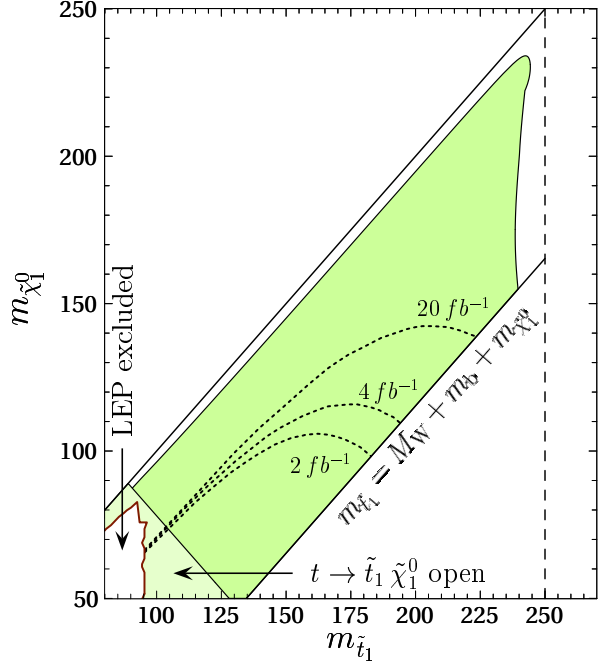, width=8.5cm}
\vspace{-2ex}
\caption{Discovery reach of linear collider with 500 fb$^{-1}$ luminosity at
$\sqrt{s} = 500$ GeV for production of light stop quarks, $e^+e^- \to
\tilde{t}_1 \, \tilde{t}_1^* \to c \bar{c} \, \neu_1 \, \neu_1$. The results
are given in the stop vs. neutralino mass plane. In the dark shaded region, a
5$\sigma$ discovery is possible. The region where $\mneu{1} > \mst$ is
inconsistent with a neutralino LSP, while for $\mst > M_{\rm W} + m_{\rm b} + 
\neu_1$ the
three-body decay $\tilde{t}_1 \to W^+ \bar{b} \neu_1$ becomes accessible and
dominant.  In the light shaded corner to the lower left, the decay of the top
quark into a light stop and neutralino is open. Also shown are the parameter
region excluded by LEP searches \cite{lep} (white area in the lower left) and
the Tevatron light stop reach \cite{Demina:1999ty} (dotted lines) for various
integrated luminosities.
}
\label{fig:cov}
\end{figure}

Based on the above results 
from the experimental simulations, the discovery reach of a 500 GeV $e^+e^-$
collider can be estimated, see Fig.~\ref{fig:cov}. 
To cover the whole parameter region, the signal efficiencies for the parameters
points in Tab.~\ref{tab:sigeff} are interpolated.
Then, the signal rates $S$ are computed by 
multiplying the efficiency $\epsilon$ obtained from the simulations with the
production cross-section for each point $(\mst,\mneu1)$.
Together with the number of background events $B$,
this yields the significance $S/\sqrt{S+B}$. The green area in the figure
corresponds to the $5\sigma$ discovery region, $S/\sqrt{S+B} > 5$.
As evident from the figure,
the ILC can find light stop quarks for mass differences down to
$\Delta m \sim {\cal O}(5 \gev)$, covering the complete stop-neutralino
co-annihilation region and beyond.

\section{STOP PARAMETER DETERMINATION}

The discovery of light stops would hint towards the possibility of
electroweak baryogenesis and may allow the co-annihilation mechanism 
to be effective.
In order to confirm this idea, the relevant
supersymmetry parameters need to be measured accurately. 
In this section, the experimental determination of
the stop parameters will be discussed.

For definiteness, a specific MSSM parameter point is chosen:
\begin{equation}
\begin{array}{rlrlrlrl}
m^2_{\rm\tilde{U}_3} &= -99^2 \gev^2,	
& A_t &= -1050 \gev,
& M_1 &= 112.6 \gev, 
& |\mu| &= 320 \gev, \\
m_{\rm\tilde{Q}_3} &= 4200 \gev,	
& \tan\beta &= 5,
& M_2 &= 225 \gev,
& \phi_\mu &= 0.2.
\end{array}
\label{eq:scen}
\end{equation}
The chosen parameters are compatible with the mechanism of electroweak
baryogenesis, generating the baryon asymmetry through the phase of $\mu$. They
correspond to a value for the dark matter relic abundance
within the WMAP bounds, $\Omega_{\rm CDM} h^2 = 0.1122$. The relic
dark matter density has been computed with the code used in Ref.~\cite{morr}.
In this scenario, the
stop and lightest neutralino masses are $m_{\tilde{t}_1} = 122.5 \gev$ and
$\mneu{1} = 107.2 \gev$, and the stop mixing angle is $\cos \theta_{\tilde{t}}
= 0.0105$, {\it i.e.} the light stop is almost completely right-chiral. The
mass difference $\Delta m = m_{\tilde{t}_1} - \mneu{1} = 15.2 \gev$ lies within
the sensitivity range of the ILC.

The measurement of $\tilde{t}_1 \tilde{t}_1^*$ production cross-section for
different beam polarization allows to extract both the mass of the light stop
and the stop mixing angle \cite{bartl97}.  Here is it assumed that 250
fb$^{-1}$ is spent each for $P(e^-)$/$P(e^+) = -80\%$/$+60\%$ and $+80\%$/$-60\%$,
where negative/positive polarization degrees indicate left-/right-handed
polarization. 
Besides the statistical errors,
the following systematic errors are taken into account: mass
measurement of the lightest neutralino $\delta \mneu{1} = 0.1 \gev$ (see
Ref.~\cite{paper} for details); determination of the polarization degree
$\delta P(e^\pm)/P(e^\pm) = 0.5\%$; error in the integrated
luminosity $\delta {\cal L} / {\cal L} = 5 \times 10^{-4}$;
theoretical uncertainty for background simulation  $\delta B / B = 0.3\%$;
stop hadronization and fragmentation 0.5--1.0\%; charm fragmentation
and tagging efficiency $\sim 0.5\%$; detector calibration $\sim 0.5\%$;
beamstrahlung uncertainty from \cite{beammoenig}.

\begin{figure}[tb]
\centering
\epsfig{file=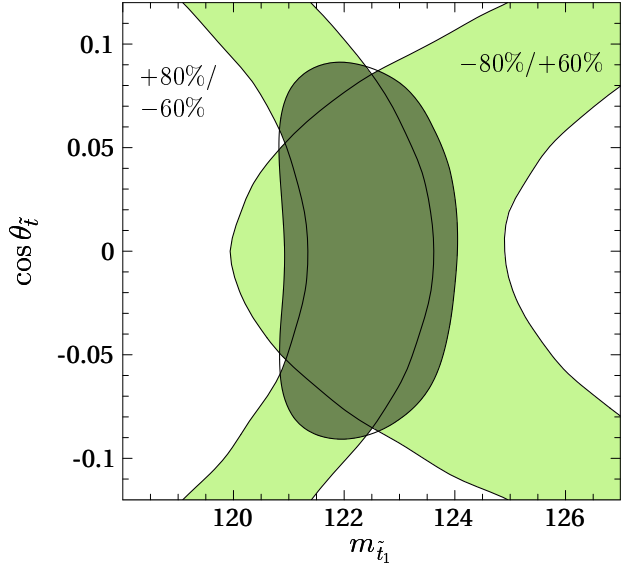, width=8.2cm, bb=20 422 306 682}
\caption{Determination of light stop mass $\mst$ and stop mixing angle
$\theta_{\tilde{t}}$ from measurements of the cross-section $\sigma(e^+e^- \to
\tilde{t}_1 \tilde{t}_1^*)$ for beam polarizations $P(e^-)$/$P(e^+)$ = 
$-$80\%/+60\% and
+80\%/$-$60\%. The plot includes statistical and systematic errors.}
\label{fig:stoppar}
\end{figure}
Each of the two cross-section measurements for $P(e^-)$/$P(e^+)$ = $-$80\%/+60\%
and +80\%/$-$60\% correspond to a band in the parameter plane of the stop mass and
mixing angle, see Fig.~\ref{fig:stoppar}. Combining the two
cross-section measurements, the stop parameter are determined to
\begin{equation}
m_{\tilde{t}_1} = (122.5 \pm 1.0) \gev, \qquad
\cos \theta_{\tilde{t}} < 0.074 \quad \Rightarrow
\sin \theta_{\tilde{t}} > 0.9972.
\end{equation}

\section{CONCLUSIONS}

Using the data from the experimental stop analysis together with estimated
errors for measurements in the neutralino/chargino sector, see
Ref.~\cite{paper}, the  expected cosmological dark matter relic density can be
computed. The mass of the heavier stop $\tilde{t}_2$ is too large to be
measured directly, but it is assumed that a limit of $m_{\tilde{t}_2} > 1000$
GeV can be set from collider searches.

\begin{figure}[tb]
\centering
\epsfig{file=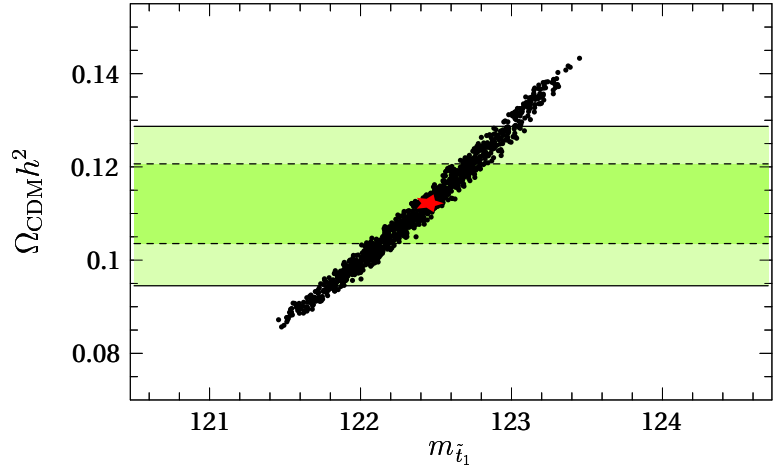, width=10.7cm}
\vspace{-2ex}
\caption{Computation of dark matter relic abundance $\Omega_{\rm CDM} h^2$
taking into account estimated experimental errors for stop, chargino, neutralino
sector measurements at the ILC. The black dots correspond to
a scan over the 1$\sigma$ ($\Delta \chi^2 \leq 1$) region allowed by the 
experimental errors, as a
function of the measured stop mass, with the red star indicating the best-fit
point. The horizontal shaded bands show the
1$\sigma$ and 2$\sigma$ constraints on the relic
density measured by WMAP.}
\label{fig:dm}
\end{figure}

All experimental errors are propagated and correlations are taken into account
by means of a $\chi^2$ analysis. The result of a scan over 100000 random points
for the scenario eq.~(\ref{eq:scen}) is shown in Fig.~\ref{fig:dm}.  The
horizontal bands depict the relic density as measured by WMAP
\cite{Spergel:2003cb}, which is at 1$\sigma$ level $0.104 < \Omega_{\rm CDM}
h^2 < 0.121$.
The collider measurements of the stop and chargino/neutralino parameters
constrain the relic density to $0.086 < \Omega_{\rm CDM} h^2 < 0.143$ at the
1$\sigma$ level, with an overall precision of the same order of magnitude,
though somewhat worse, than the direct WMAP determination. 
The uncertainty in the computation of the dark matter relic density from
ILC precision measurements is dominated by the measuremenst of the
lightest stop mass.

\begin{acknowledgments}
The authors are grateful to P.~Bechtle, S.~Mrenna and T.~Kuhl for practical
advice and to M.~Schmitt for very useful discussions.
\end{acknowledgments}



\begin{thebibliography}{99} 

\bibitem{Spergel:2003cb}
D.~N.~Spergel {\it et al.}  [WMAP Collaboration],
Astrophys.\ J.\ Suppl.\  {\bf 148}, 175 (2003);
M.~Tegmark {\it et al.}  [SDSS Collaboration],
Phys.\ Rev.\ D {\bf 69}, 103501 (2004).

\bibitem{Balazs:2004bu}
C.~Bal\'azs, M.~Carena and C.~E.~M.~Wagner,
Phys.\ Rev.\ D {\bf 70}, 015007 (2004).

\bibitem{Demina:1999ty}
R.~Demina, J.~D.~Lykken, K.~T.~Matchev and A.~Nomerotski,
Phys.\ Rev.\ D {\bf 62}, 035011 (2000).

\bibitem{stopsnew}
  A.~Finch, H.~Nowak and A.~Sopczak,
in {\it Prec. of the International Europhysics Conference on High-Energy Physics (HEP 2003), Aachen, Germany, 17-23 Jul 2003}
[LC Note LC-PHSM-2003-075].

\bibitem{pythia}
  T.~Sj\"ostrand {\it et al.},
  Comput.\ Phys.\ Commun.\  {\bf 135}, 238 (2001).
  
\bibitem{circe}
T.~Ohl,
Comput.\ Phys.\ Commun.\  {\bf 101} (1997) 269.

\bibitem{simdet}
  M.~Pohl and H.~J.~Schreiber,
  hep-ex/0206009.

\bibitem{slep}
A.~Freitas, D.~J.~Miller and P.~M.~Zerwas,
Eur.\ Phys.\ J.\ C {\bf 21} (2001) 361;
A.~Freitas, A.~von Manteuffel and P.~M.~Zerwas,
Eur.\ Phys.\ J.\ C {\bf 34} (2004) 487.

\bibitem{grace}
  F.~Yuasa {\it et al.},
  Prog.\ Theor.\ Phys.\ Suppl.\  {\bf 138}, 18 (2000).

\bibitem{comphep}
  E.~Boos {\it et al.}  [CompHEP Collaboration],
  Nucl.\ Instrum.\ Meth.\ A {\bf 534}, 250 (2004).

\bibitem{kuhlhiggs}
T.~Kuhl,
  in {\it Proc. of the International Conference on Linear Colliders (LCWS 04), 
  Paris, France, 19-24 Apr 2004};
T.~Kuhl, LC Note in preparation.
  
\bibitem{lep}
LEP2 SUSY Working Group, ALEPH, DELPHI, L3 and OPAL experiments, note
LEPSUSYWG/04-02.1. 

\bibitem{morr}
  C.~Bal\'azs, M.~Carena, A.~Menon, D.~E.~Morrissey and C.~E.~M.~Wagner,
  Phys.\ Rev.\ D {\bf 71}, 075002 (2005).

\bibitem{bartl97}
A.~Bartl, H.~Eberl, S.~Kraml, W.~Majerotto, W.~Porod and A.~Sopczak,
  Z.\ Phys.\ C {\bf 76}, 549 (1997).

\bibitem{paper}
M.~Carena, A.~Finch, A.~Freitas, C.~Milst\'ene, H.~Nowak, A.~Sopczak,
hep-ph/0508152.

\bibitem{beammoenig}
  K.~M\"onig,
in {\it Proc. of the 2nd ECFA/DESY LC Study (1998-2001)}, p.   
1353-1361 [LC Note LC-PHSM-2000-060].

\end{thebibliography}
\end{document}